\documentclass[10pt,times,twocolumn]{article}

\pdfoutput=1

\usepackage[all]{xy}
\usepackage{color,url}
\usepackage{subfigure}
\usepackage{bm,amsmath}
\usepackage{marvosym}
\usepackage{graphicx}
\usepackage{wrapfig}
\usepackage{epsf}
\usepackage{times,mathptm}
\usepackage{url}

\begin{document}

\title{\Large \bf Voltage Collapse and ODE Approach to Power Flows:\\
Analysis of a Feeder Line with Static Disorder in Consumption/Production}

%{\color{red} !!!!! Authors are needed only for the final submission - hide before submitting}
\author {{\bf Michael Chertkov} $^{(1)}$, {\bf Scott Backhaus} $^{(2)}$, {\bf Konstantin Turtisyn} $^{(3)}$,\\ {\bf Vladimir Chernyak} $^{(4,1)}$ and {\bf Vladimir Lebedev} $^{(5)}$\\
Center for Nonlinear Studies and Theoretical Division $^{(1)}$ and \\
Material Physics and Applications Division $^{(2)}$, LANL, Los Alamos, NM 87545, USA\\
Mechanical Engineering, MIT, Cambridge, MA 02139, USA $^{(3)}$\\
Department of Chemistry, Wayne State University, Detroit, MI 48202, USA $^{(4)}$\\
Landau Institute for Theoretical Physics, Kosygina 2, Moscow 119334, Russia $^{(5)}$
}
%LA-UR 11-03433

\maketitle

\begin{center}
{\bf \large Abstract}
\end{center}
{\it We consider a model of a distribution feeder connecting multiple loads to the sub-station. Voltage is controlled directly at the head of the line (sub-station), however, voltage anywhere further down the line is subject to fluctuations, caused by irregularities of real and reactive distributed power consumption/generation. The lack of a direct control of voltage along the line may result in the voltage instability,  also called voltage collapse - phenomenon well known and documented in the power engineering literature. Motivated by emerging photo-voltaic technology, which brings a new source of renewable generation but also contributes significant increase in power flow fluctuations,  we reexamine the phenomenon of voltage stability and collapse. In the limit where the number of consumers is large and spatial variations in power flows are smooth functions of position along the feeder, we derive a set of the power flow Ordinary Differential Equations (ODE),  verify phenomenon of voltage collapse, and study the effect of disorder and irregularity in injection and consumption on the voltage profile by simulating the stochastic ODE. We observe that disorder leads to nonlinear amplification of the voltage variations at the end of the line as the point of voltage collapse is approached. We also find that the disorder, when correlated on a scale sufficiently small compared to the length of the line,  self-averages, i.e. the voltage profile remains spatially smooth for any individual realization of the disorder and is correlated only at scales comparable to the length of the line. Finally,  we explain why the integrated effect of disorder on the voltage at the end of the line cannot be described within a naive one-generator-one-load model.
}

\vspace{0.5cm}

\section{Introduction}
\label{sec:Intro}

Voltage collapse/instability is a severe disturbance but also one of the most interesting and most discussed nonlinear phenomenon in power engineering, with the bibliography \cite{98AL} published in 1998 containing 308 references and a review in 2000 \cite{VanCutsem2000} containing 132 references. According to \cite{98AL}, the first paper related to voltage instability appeared in \cite{Weedy1968}, and the first criteria for detecting the point of voltage collapse was proposed in \cite{Venikov1975}. A number of comprehensive books \cite{Venikov1977,Taylor1994,Cutsem1998} and a book chapter in \cite{Kundur1993} are written on the subject.

In this manuscript, we give an applied math and statistical physics prospective on description of the voltage collapse phenomenon in the spatially continuous (ODE) limit, and we also begin to investigate effects of spatial disorder in power injection/generation on the voltage instability.  To the best of our knowledge, an ODE description of power flows was discussed in the power-engineering literature only in the context of the electro-mechanical waves \cite{Parashar2004,Thorp1998}, and not to describe voltage collapse and instability. We are also not aware of any discussion in the literature of the effect of spatial disorder, and resulting structural stability, in power consumption along a power line \footnote{Note, that the effect of spatio-temporal noise in consumption on the voltage collapse in a transmission system was already discussed in \cite{DeMarco1987}.}.

A natural question that should be answered before the development of ODE models is whether the continuous models are useful for modeling discrete systems like  the power grid. Similar questions have been answered multiple times in the field of statistical mechanics which aims at establishing connections between microscopic models of various materials (for instance gas of interacting particles) and their thermodynamic properties. The main lesson learned from decades of research in this area is that all details of microstructure are in fact not that important when one is focused on describing the large-scale properties of the system. The starting point of our approach is the observation that power networks share many properties with statistical mechanics systems. Huge number of individual loads contribute to the power flow; however, modification, inclusion or removal of any individual load will not lead to any visible macroscopic change. This general macroscopic approach motivates us to ask the question: is a coarse-grained, continuous ODE description a proper tool to describe macroscopic phenomena in power grids?

To answer this fundamental question we extended the ODE analysis of the voltage collapse/instability to an inhomogeneous/disordered case. Within the ODE description, we consider effects of temporally frozen but spatially varying irregularity and disorder in real and reactive consumption and generation on the voltage profile in a representative feeder line of a power distribution system. \footnote{Note,  however,  that the developed formalism is generic and as such it as well applies to linear segments of transmission lines,  which will also be discussed briefly in the following.} Assuming that the correlation scale of the disorder is smaller than the length of the feeder (but larger than the inter-distance between neighboring nodes), we study the resulting stochastic power plow (PF) ODEs in simulations. Our main findings consists in the following three observations:
\begin{itemize}
\item The effect of disorder on the voltage profile,  and specifically in voltage variations at the end of the line, is amplified in a significant way with the parameters of the base solution (uniform component of consumption along the line) as the point of the voltage collapse is approached. The probability of observing voltage collapse is increased in a nonlinear fashion in the presence of disorder. %{\bf I DO NOT UNDERSTAND THIS SENTENCE This nonlinear effect is associated with a nonlinear (activation type) increase in probability for disorder to result, once added to the uniform consumption/generation pattern, in the loss of a feasible solution of the Power Flow ODE.}

\item Any individual configuration of the disorder results in a smooth spatial profile of voltage and power flows, correlated at the scale roughly associated with the length of the feeder line, and which is certainly much longer then the correlation scale of the disorder. We attribute this self-averaging effect to the fact that solutions for the power flows,  but especially for the voltage, emerge as spatial integrals. The observed smoothness of the solution implicitly justifies using the reduced ODE approach in the power systems.

\item The effect of disorder on the voltage at the end of the long feeder line cannot be described properly by a reduced one-generator-one-load model,  i.e. the effects revealed by the stochastic ODE description cannot realized in a lumped model by integrating the injection and consumption of real and reactive powers along the line.
\end{itemize}

The material in the manuscript is organized as follows. Section \ref{sec:TechIntro} contains a technical introduction resolving in the ODE formulation of power flows for linear segments of the grid.
Section \ref{sec:static} analyzes phenomenon of voltage collapse and instability within the ODE approach, discussing both a linear element of a transmission system and a linear feeder line of a distribution system. Section \ref{sec:str-stab} is devoted of analysis of a structural stability of a feeder line subjected to static variations in consumption and injection along the line which are correlated on a short length scale. Section \ref{sec:conclusions} is reserved for conclusions and brief discussions of the path forward.

\section{Power Flows in Discrete and Continuous (ODE) Formulations}
\label{sec:TechIntro}

Static Power Flow (PF) equations over a power graph/network, ${\cal G}=({\cal V},{\cal E})$, are reformulations of the AC Kirchoff laws for currents and voltages defined at any vertex of the graph, $a\in {\cal V}$:
\begin{eqnarray}
&&
0=\tilde{p}_a- \sum_{b\sim a}
v_a v_b\sin(\theta_a-\theta_b)
\frac{X_{ab}}{R_{ab}^2+X_{ab}^2}\nonumber\\ &&
-\sum_{b\sim a}
\left(v_a^2-v_a v_b\cos(\theta_a-\theta_b)\right)
\frac{R_{ab}}{R_{ab}^2+X_{ab}^2},\label{PF_theta}\\
&& 0=\tilde{q}_a-
\sum_{b\sim a}\!\left(v_a^2\!-\! v_a v_b \cos(\theta_a-\theta_b)\right)
\!\frac{X_{ab}}{R_{ab}^2+X_{ab}^2}\nonumber\\ &&
+\sum_{b\sim a} v_a v_b\sin(\theta_a-\theta_b)
\frac{R_{ab}}{R_{ab}^2+X_{ab}^2},\label{PF_v}
\end{eqnarray}
where $\tilde{p}_a,\tilde{q}_a,v_a,\theta_a$ are real and reactive power injections (negative for consumption), potential, and phase at the node $a\in{\cal V}$, respectively.  Here, $b\sim a$ indicates that $b$ and $a$ are graph-neighbors connected by an edge $(a,b)$ of ${\cal E}$ characterized by resistance $R_{ab}$ and inductance $X_{ab}$. In transmission (high voltage) lines, resistance is significantly smaller than reactance ($R_{ab}\ll X_{ab}$), while in the distribution (low voltage) networks resistance and reactance are usually of the same order.

Relations between the electric potential (voltage and phase) and powers (real and reactive) are universal, however, what constitutes input or output vary for nodes of different types. Standard generator nodes are kept at constant voltage (assumed unity in the rescaled units), and thus their input pair is real power and voltage, i.e. they are $(p,v)$ nodes. Regular consumer nodes draw prescribed powers,  therefore these are $(p,q)$ nodes with real and reactive powers being the two input parameters. The largest generator of a system, or a node connecting the system to a bigger power network, is taken as a phase reference where both voltage and phase are assumed fixed, i.e. a $(v,\theta)$ node.  Summarizing, the power flow equations Eq.~(\ref{PF_theta},\ref{PF_v}) constitute a system of algebraic equations expressing unknown (output) variables,  which are real and reactive powers at the super node, reactive powers and phases at the generator nodes, and phases and voltages at the consumer nodes,  via the set of known (input) node variables.

Note that the $(p,q)$ model with constant and time-steady  $p$ and $q$ is a relatively crude approximation. In reality,  $p$ and $q$ do depend on instantaneous voltage at the load $v$, the local rate of phase variation $\dot{\theta}$ (i.e. on the frequency), and on the rate of voltage variation $\dot{v}$. Modeling the so-called static [$(p,q)(v)$], and dynamic [$(p,q)(\dot{\theta},\dot{v})$] loads are classic and also difficult subjects in power engineering. See a (relatively) early discussions of the load modeling in \cite{Dobson1989a,Hill1993,Davy1997} and \cite{Lesieutre2005,PNNL-16916} for description of the current status quo and problems  in the reduced but (still multi-parametric) composite modeling of loads within Western Electricity Coordinating Council (WECC) \cite{WECC} presented as a part of the GE Positive Sequence Load Flow (PSLF) software \cite{PSLF}. The main difficulty in the microscopic modeling of loads comes from the uncertainty and variability of loads over time. Indeed, an aggregated load representing a node of the grid may consist of a very special combination of nonlinear and dynamic elements(like rotors) and nonlinear and linear static elements (like resistors or inductances). On the time scale of interest here (seconds or minutes), an aggregated load load model is not expected to change in time, i.e. it is quenched. The situation poses the question of ``learning" the nonlinear and dynamic (but reduced) models of loads.

It is common in a transmission network to have a linear segment connecting two generators with multiple loads drawing power from the line in between. Feeder lines are the most typical elements of distribution networks where strong connection to the transmission network resides at the head of the line, and the line ends at the last load of the linear segment. Assuming that the loads are distributed uniformly and frequently along the line so that the inter-load spacing $a$ is sufficiently small ($a\ll L$), the voltages and phases can be expanded in a Taylor series in $a$,
$v_{n+1}=v_n+(\partial_z v)_n a+(\partial^2_z v)_n a^2/2+O(a^3)$, $\theta_{n+1}=\theta_n+(\partial_z \theta)_n a+(\partial^2_z \theta)_n a^2/2+O(a^3)$.  Substituting the expansions into Eqs.~(\ref{PF_theta},\ref{PF_v}) and keeping only the leading $O(a^2)$ terms, one arrives at
\begin{eqnarray}
&& \hspace{-0.5cm}
0%=\frac{\tilde{p}-p_{el}}{a}
=p_{mech}+\beta\partial_r \left(v^2\partial_z\theta\right)+g v\left(\partial_z^2 v-v \left(\partial_z \theta\right)^2\right),\label{PF_theta_cont}\\
&&
\hspace{-0.5cm}
0%=\frac{\tilde{q}-q_{el}}{a}
=q_{mech}+\beta v\left(\partial_z^2 v-v \left(\partial_z \theta\right)^2\right)-g\partial_z \left( v^2\partial_z\theta\right),\label{PF_v_cont}
\end{eqnarray}
where  $p_{mech}=\tilde{p}/a, q_{mech}=\tilde{q}/a$ are real and reactive power densities \footnote{The subscript ``mech" in $p$ and $q$ indicates that the powers are mechanical powers at the buses. In the following we will skip the subscript to lighten notations.}, $g=a R/(R^2+X^2)$ and $\beta=a X/(R^2+X^2)$ are conductance and susceptance measured in the units of the inter-load spacing, $a$. Note that the two second-order ODEs~(\ref{PF_theta_cont},\ref{PF_v_cont}) can be restated as the following four first-order ODEs (written at $\beta=1$ for simplicity of notations )
\begin{eqnarray}
\partial_z\left(\begin{array}{c} \theta\\ v \\ s \\ w \end{array}\right)=
\left(\begin{array}{c} \frac{s}{v^2} \\ w \\ -\frac{p+g q}{1+g^2} \\
\frac{s^2}{v^3}-\frac{\beta p+q}{(1+g^2)v}\end{array}\right).
\label{four_ode}
\end{eqnarray}

Flows of real and reactive powers
\begin{eqnarray}
P=-\beta v^2\partial_z\theta -g v\partial_r v,\quad Q=-\beta v \partial_z v+g v^2\partial_z\theta,
\label{PQ_cont}
\end{eqnarray}
are generally dependent on the position along the line $z$. These two objects, which appears on the rhs of Eqs.~(\ref{PF_theta_cont}), are important characteristics of the flows. (See Appendix \ref{sec:DistFlow} for more details.)

For a transmission line,  the four natural boundary conditions (for two second order differential equations) are known voltage at the two ends of the line [$v(0)=v(L)=1$], zero (reference) phase at the head of the line [$\theta(0)=0$], and a fixed value of the real power flow at the head of the line [$P(0)$]. For a distribution line, the four natural boundary conditions become: known voltage at the head of the line [$v(0)=1$], zero (reference) phase at the head of the line [$\theta(0)=0$], and zero flow (flux) for both real and reactive powers at the end of the line [$P(L)=Q(L)=0$].

Note that the PF equations, in discrete (\ref{PF_theta},\ref{PF_v}) or continuous (\ref{PF_theta_cont},\ref{PF_v_cont}) forms, may show multiple solutions or have no solutions.
It is also important to mention,  even though we do not discuss in the manuscript, that the ODE-based static power flow framework naturally fits in a more general dynamic framework, accounting for dynamics and control of generators and loads.

\section{Static Analysis of the Continuous Model in the Spatially Uniform Case}
\label{sec:static}

Our first step is to verify that transition to the ODE description of a linear segment (for a transmission network or a feeder line) preserves the phenomenon of the voltage collapse/instability  observed in discrete power systems.

%{\color{red} ... so far it contains the full package from the ``prl2"-file it may be reduced somehow ...}

\begin{figure}[t]
\centering \caption{Lowest voltage along the $[0;L]$ segment vs the length of the segment for fixed values of load consumption ($p=-5$ and $q=-1$), conductance ($g=0.5$),  and susceptance ($\beta=1$).  The value of the real power injected in the line at $L=0$ is varied. Dashed curves show the respective voltage profiles along the critical segment, $z\in[0;L_c]$. }
\includegraphics[width=0.5\textwidth]{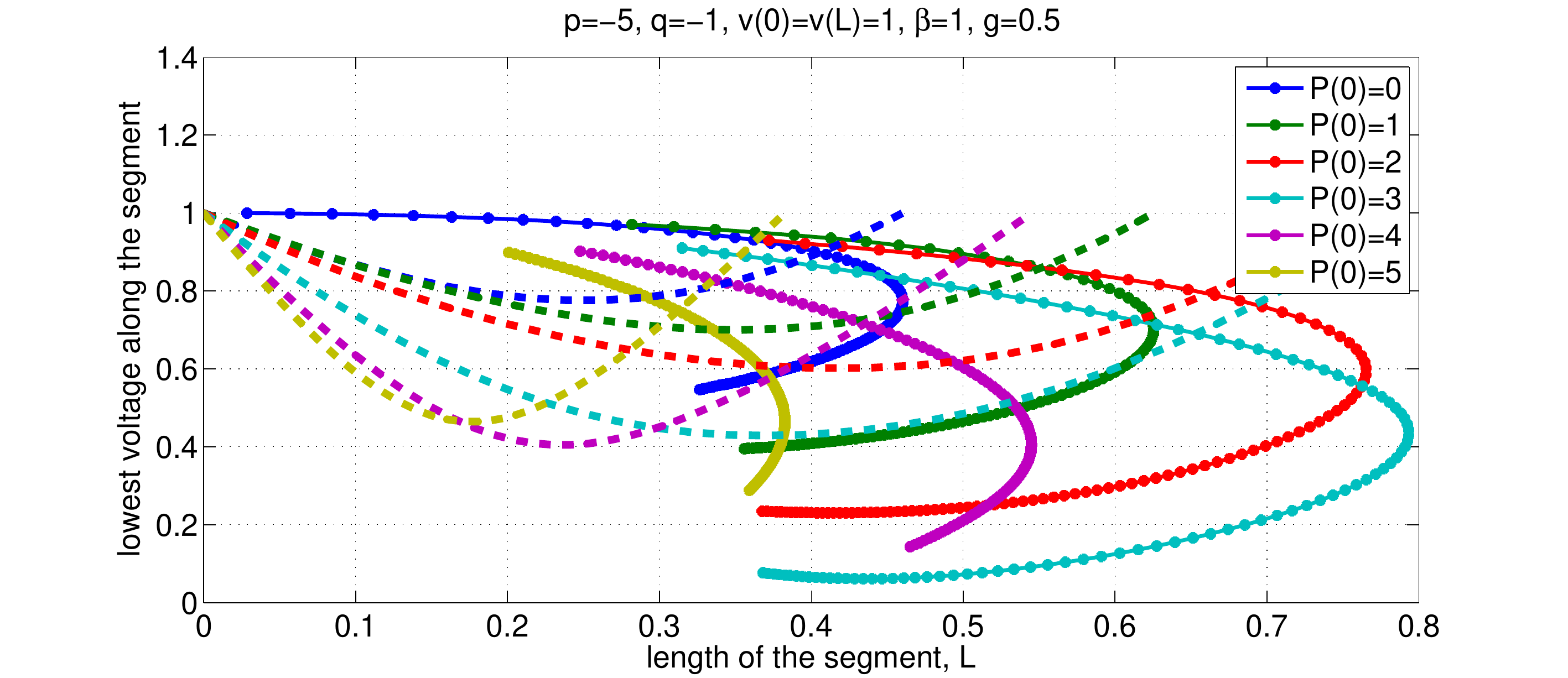}
\label{fig:shooting1}
% figure(1) of test_shooting.m
\end{figure}

\begin{figure}
\centering \caption{Profile of real (solid) and reactive (dashed) powers flowing along the line.
The curves correspond to the critical segments with the parameters (and color-coding) from Fig.~(\ref{fig:shooting1}).
 }
\includegraphics[width=0.5\textwidth]{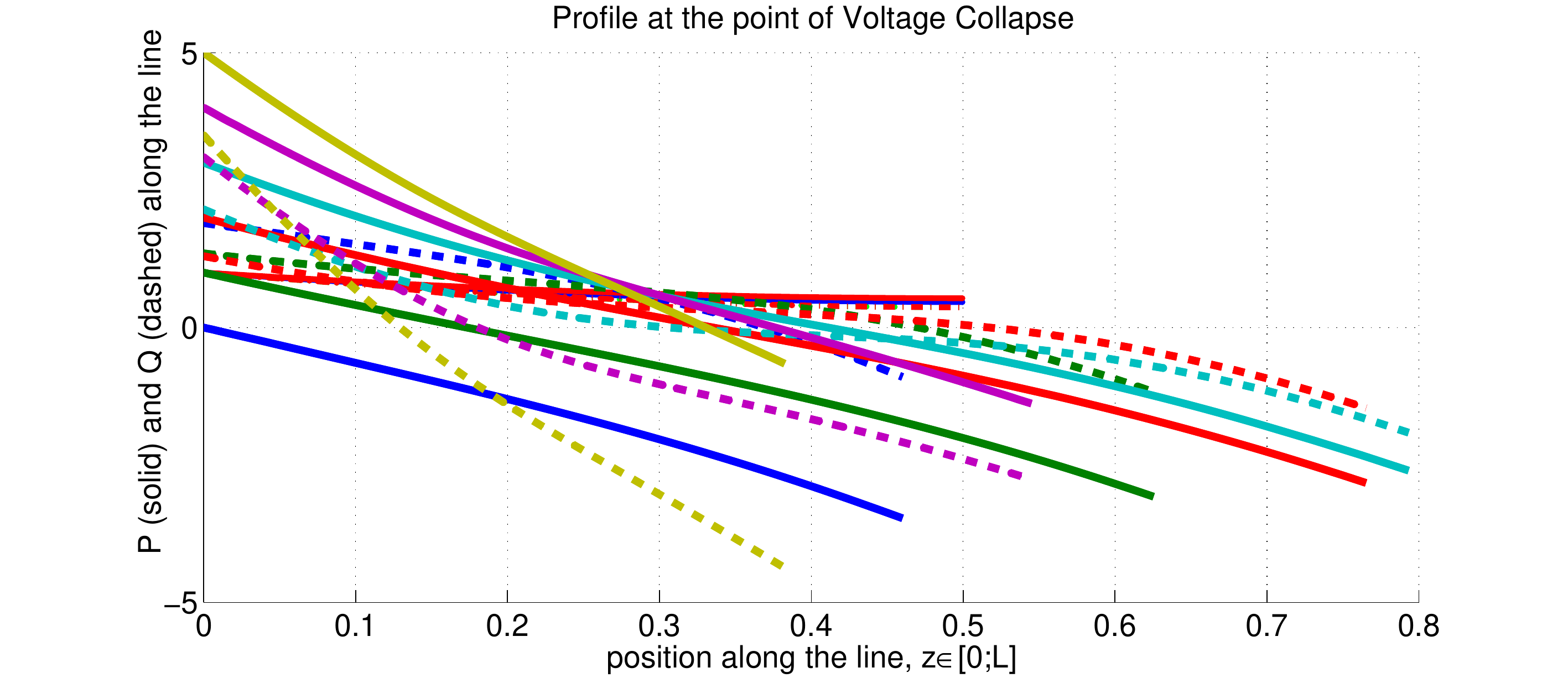}
\label{fig:shooting2}
% figure(2) of test_shooting.m
\end{figure}

We consider the ODE PF model (\ref{PF_theta_cont},\ref{PF_v_cont}) of a transmission line with constant $p,\;q,\;g,$ and $\beta$ homogeneously distributed along the line $z\in[0;L]$ with voltage fixed at the both ends of the line $v(0)=v(L)=1$. We also fix the value of a real power $P(z=0)$ injected at the head of the line, and we study the dependence of the voltage and phase profile along the line as a function of $L,\;\nabla\theta(0),\;p$ and $q$. The set of two ODEs allows analytic solution (in quadratures) only in the special (but unphysical) case of $p=0$,  therefore we largely rely here on a straightforward numerical study of Eqs. (\ref{PF_theta_cont},\ref{PF_v_cont}).  Simulations are performed using the shooting method.

Typical results of direct illustrating phenomenon of voltage collapse are shown in Figs.~(\ref{fig:shooting1},\ref{fig:shooting2}). In this static interpretation, voltage collapse consists in the disappearance of a valid solution for sufficiently long lines of length $L>L_c$, where thus $L_c$ is the critical value corresponding to the point of voltage collapse. For $L<L_c$, valid solutions appear in pairs, one stable and one unstable correspondoing to high and low voltage, respectively. For $L<L_c$, these two solutions vary with $L$, and they merge at the critical point $L=L_c$.  Modifying $g/\beta$ does not change the qualitative shape of the curves in the Figs.~(\ref{fig:shooting1},\ref{fig:shooting2}). We note a non-monotonic dependence of $L_c$ and $v_c$ (voltage at the point of collapse) on $P(0)$.

Main conclusion one draws from  Figs.~(\ref{fig:shooting1},\ref{fig:shooting2}) and related analysis is that the ``nose shape" behavior is a universal phenomenon. The stable solution corresponds to higher throughput (of both real and reactive) power over the line. On the quantitative side, position of the nose (both in terms of the critical length of the line and the critical voltage) is a non monotonic function of the line parameters: the power consumed along the line, the power dissipated within the line, and power transferred through the line. The line is the longest in the symmetric situation of the zero throughput.

\begin{figure}[t]
\centering \caption{Dependence of the voltage on the density of real power consumption along the feeder for $L=0.5$ and $\partial_z\theta(L)=\partial_z v(L)=0$. Doted and dashed curves correspond to stable and unstable branches respectively.}
\includegraphics[width=0.5\textwidth]{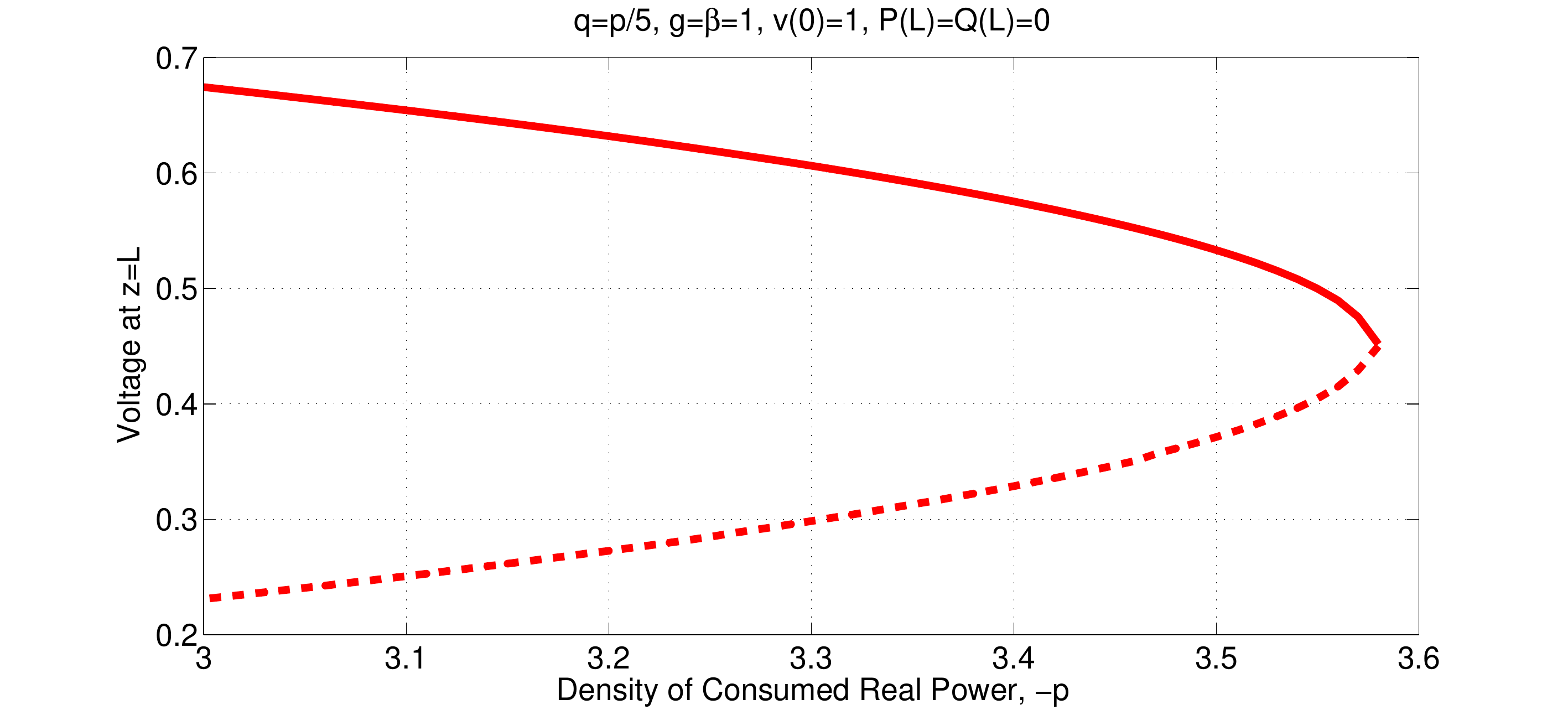}
\label{fig:four1}
% figure(1) of test_four_feeder.m
\end{figure}

\begin{figure}
\centering \caption{Dependence of the real and reactive power at the head of the line on the density of real power consumption, complementing description of Fig.~(\ref{fig:four1}). Solid and dashed curves show stable and unstable branches respectively.}
\includegraphics[width=0.5\textwidth]{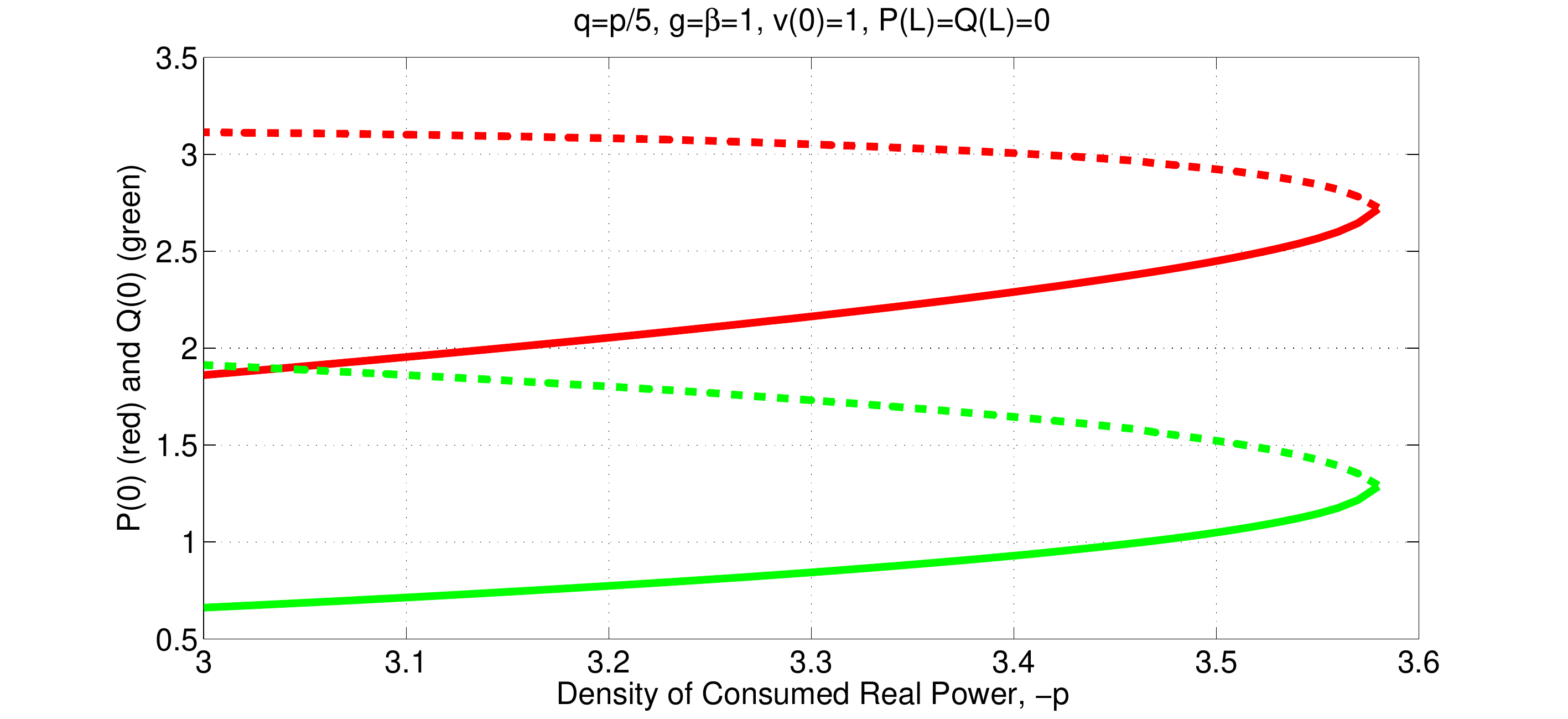}
\label{fig:four2}
% figure(2) of test_four_feeder.m
\end{figure}

Next, we analyze a feeder line of fixed length $L$ where the end of the line is hanging free and the amount of real and reactive powers injected at the beginning of the feeder is adjusted to the following four boundary conditions, $v(0)=1,\theta(0)=0,P(L)=Q(L)=0$. The boundary value problem is studied with the help of the bvp4c Matlab solver.  The results are shown in the Figs.~(\ref{fig:four1},\ref{fig:four2}) and describe the ``nose shape" for a feeder of the fixed length.

Here we conclude that the two solutions,  one stable and another unstable, are observed for a feeder of a fixed length when the amount of real and reactive power consumed (homogeneously) through the feeder is smaller than the critical value, corresponding to the nose of the curve in Fig.~(\ref{fig:four1}). We also observe the power injected at the head of the line is smaller for the stable solution than for the unstable solution, confirming the intuition that the choice of the stable solution is consistent with the minimal energy/power principe.

\section{Structural Stability Analysis}
\label{sec:str-stab}

\begin{figure}[t]
\centering \caption{Voltage profile (twenty trials per case/color) for three base configurations of the feeder line and  added Gaussian finite-correlated disorder (in both $p$ and $q$) of the same strength. (Red blue and green corresponds to $p=-2.5,-3$ and $-3.5$ respectively.) Low and Upper branches (solid and dashed) correspond to stable and unstable solutions. See text for further details.}
\includegraphics[width=0.5\textwidth]{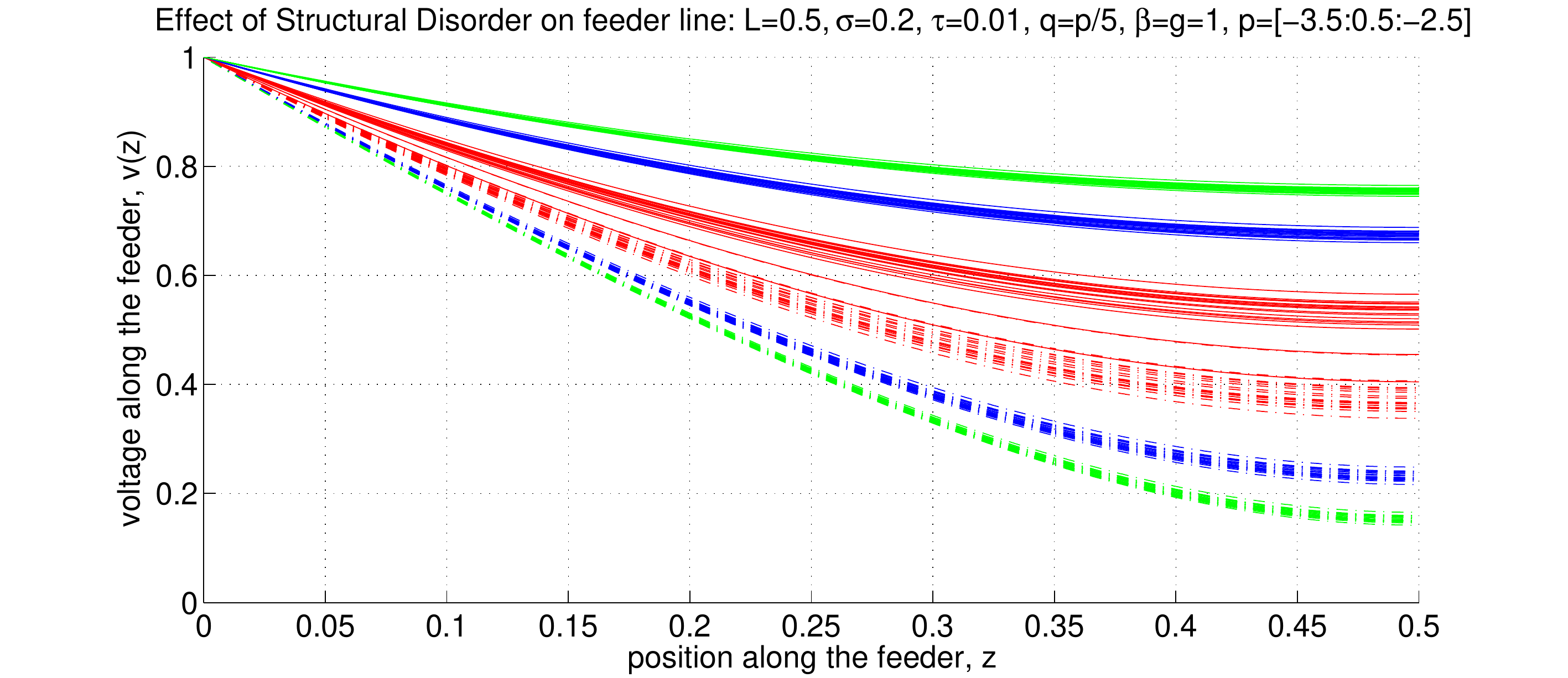}
\label{fig:struct_disorder}
% figure(2) of test_four_feeder_random_comparison.m
\end{figure}

\begin{figure}
\centering \caption{Multiple (twenty trials per case/color) configurations of the disorder resulted in voltage profiles shown in Fig.~(\ref{fig:struct_disorder}).}
\includegraphics[width=0.5\textwidth]{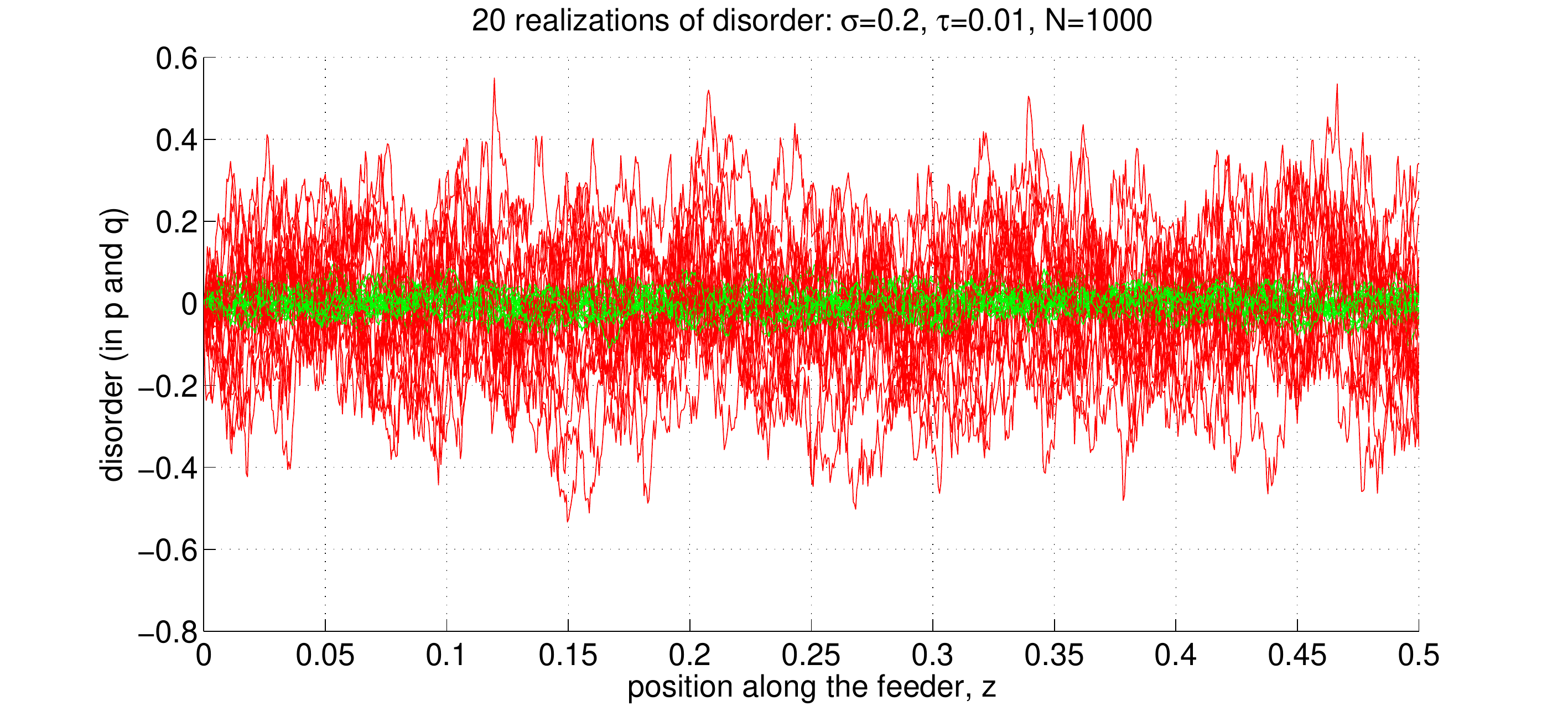}
\label{fig:disorder_config}
% figure(1) of test_four_feeder_random_comparison.m
\end{figure}

\begin{figure}
\centering \caption{Comparison of the actual disorder effect (one realization from the ensemble shown in Fig.~(\ref{fig:disorder_config})) on the voltage profile (star-line) with an aggregated effect (dash-line),  where in the latter case $p$ and $q$ along the line are kept uniform and correspondent to $p_0$ and $q_0$ with added small values,  $\int_0^r \delta_p(z')dz'/L$ and $\int_0^z \delta_q(z')dz'/L$ respectively. The green is not seen behind red - the curves virtually coincide.}
\includegraphics[width=0.5\textwidth]{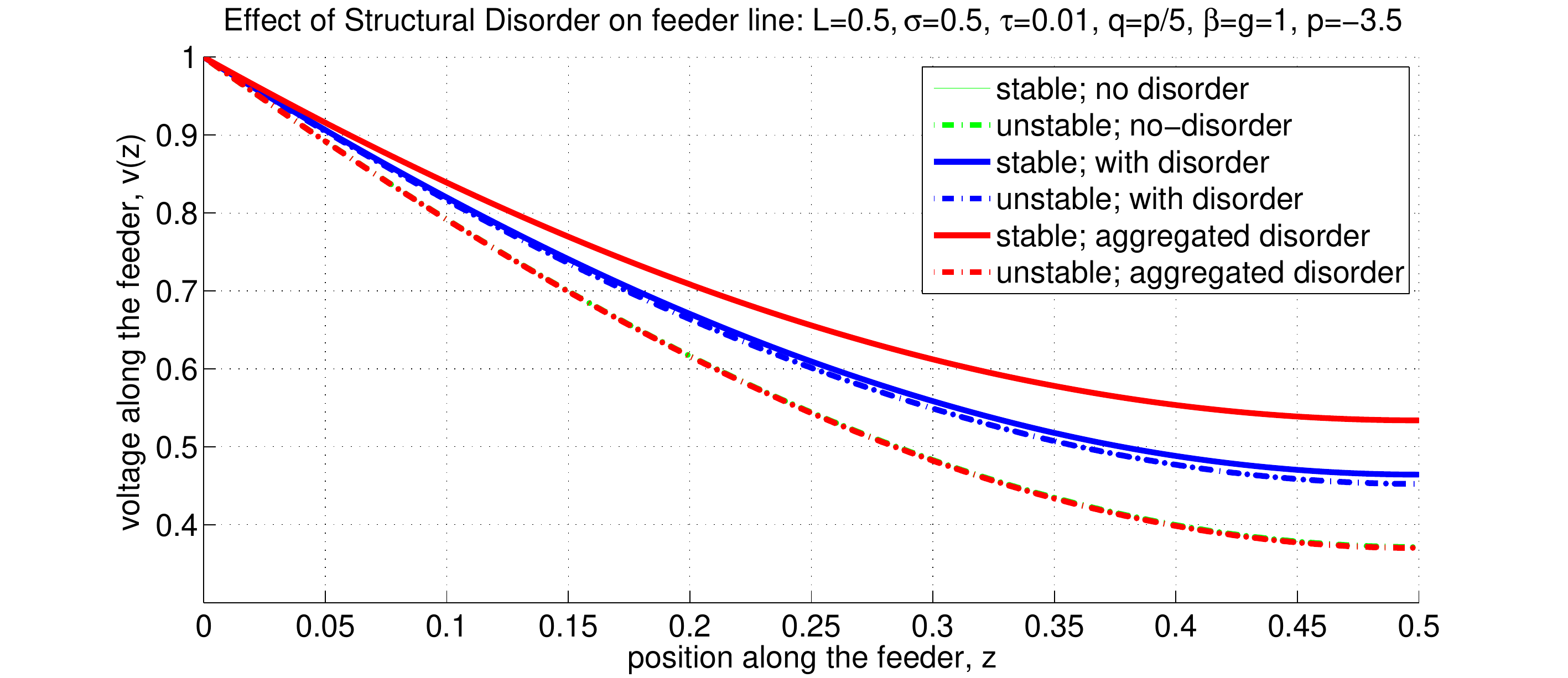}
\label{fig:disorder3a}
% figure(2) of test_four_feeder_random3.m
\end{figure}

\begin{figure}
\centering \caption{Configuration of disorder correspondent to the profiles shown in Fig.~(\ref{fig:disorder3a}).}
\includegraphics[width=0.5\textwidth]{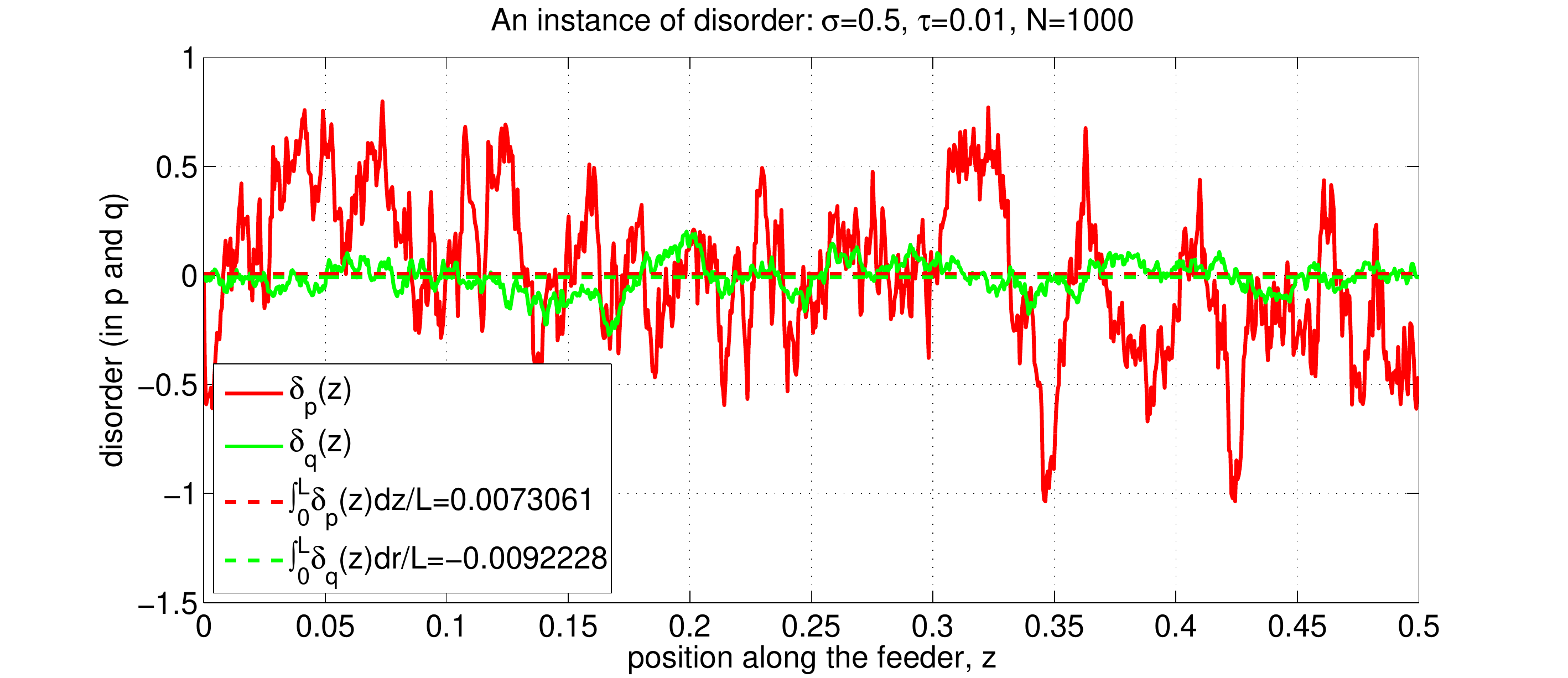}
\label{fig:disorder3b}
% figure(1) of test_four_feeder_random3.m
\end{figure}

Structural stability pertains to modification of the typical``nose" curve shown in Fig.~(\ref{fig:four1}) because of structural disorder  in the  $p$ and $q$, i.e. inhomogeneity in the $p,q$ distribution along the line.

One important phenomenon, illustrated in Fig.~(\ref{fig:struct_disorder}), is the amplification of the effect of disorder when approaching the point of voltage collapse. We tested three sample cases from Figs.~(\ref{fig:four1},\ref{fig:four2}) corresponding to an average density of consumption $p_0=-2.5;-3;-3.5$ with disorder in $p,q$ added. The disorder, $\delta_p(z)$ and $\delta_q(z)$ where $p(z)=p_0+\delta_p(z)$ and $q(z)=q_0+\delta_q(z)$, is modeled as zero-mean, colored Gaussian noise characterized by $\langle \delta_p(z)\delta_p(z')\rangle=2\sigma^2\exp(-|z-z'|/\tau)$ and $\langle \delta_q(z)\delta_q(z')\rangle=2(\sigma/5)^2\exp(-|z-z'|/\tau)$, where $\sigma=0.5,\tau=0.01$.  Multiple realizations of the disorder are shown superimposed in Fig.~(\ref{fig:disorder_config}).

As illustrated in Figs.~(\ref{fig:disorder3a},\ref{fig:disorder3b}), the shift in the voltage profile due to disorder is not associated with a simple renormalization of the cumulative loads $\int p(z')dz'/L$ and $\int q(z') dz'/L$ along the line. The fact that the curves without disorder and uniformly distributed disorder coincide in Fig.~(\ref{fig:disorder3a}) is incidental. The order of curves and relative difference between them may vary from one disorder configuration to another,  but in general differences between three type of curves (no-disorder, disorder, and uniformly distributed disorder) are of the same order. In some cases the disorder may significantly degrade the situation but in others it may improve it. In some rare cases the correction added by disorder was super-critical thus resulting in the algorithm failing to find any solution.

Two main conclusions drawn from these simulations are as follows.
\begin{itemize}
\item In spite of the the fact that the amount of disorder in $p$ and $q$ added was identical in all the three cases, the spread of voltage was nonlinearly amplified as criticality is approached (the point of voltage collapse). The nonlinear amplification is associated with the fact that the closer the system is to the point of the voltage collapse, the easier it becomes for disorder to push the system even closer towards instability or possibly overshoot into the unstable regime. Since the behavior is highly nonlinear in the vicinity of the point of collapse, the effect of amplification is nonlinear too.

\item The effect of disorder on the voltage profile self-averages: the voltage profile decreases monotonically even after the sub-critical configuration of the disorder is added (when the system still has a dynamically stable solution). Moreover, we have also observed that the overall effect of the disorder cannot be explained simply as a renormalization of the total power consumed along the line. Adding disorder may push the system to criticality or relieve it somewhat.  The self-averaging feature of the resulting voltage profile is associated with the nature of power flow equations and boundary conditions. Indeed, the DistFlow ODE representation (explained in the Appendix) makes this clear.  Eqs.~(\ref{QP_cont}) are integrated backwards in $z$ from $L$, substituted into Eq.~(\ref{v_eq}), and integrated forward in $z$.  This double-integral structure results in smoothing of the input/disorder in the voltage profile/output. It is also obvious that the resulting profile is not a simple integral of the disorder, but rather a nontrivial convolution of the disorder with a kernel that is nonuniform along the feeder line.
\end{itemize}

\section{Conclusions and Path Forward}
\label{sec:conclusions}

In this manuscript we developed an ODE approach to static analysis of voltage instability collapse in linear segments of transmission and distribution power systems. The technique allows the use of powerful ODEs/PDEs tools of applied mathematics and statistical physics for deriving a reduced, coarse-grained description of power flows and voltage variations along these lines.  The technique was validated on examples of lines with uniform and homogenenous distributions of consumption/generation of real and reactive power. The PF ODEs shows voltage collapse and instability, similar to these observed in finite systems. Then we used the technique to study effects of variations and disorder in consumption/generation on voltage profile.  We observe that the effect of disorder, in particular on the value of voltage at the end of the feeder line, is amplified in proximity of the voltage collapse. Moreover, we observe that the resulting voltage profile remains smooth even under action of a short-correlated disorder, which justifies the use of the reduced ODE approach for power system modeling.

This manuscript is our first paper reporting results of the ODE analysis of voltage stability in power systems.  We envision extending this research along the following (and possibly many other) directions.
\begin{itemize}
\item Introduction of dynamic (deterministic and stochastic) modeling of load and generation and studying dynamic phenomena underlying the static analysis of this paper. We will be developing practical tools for analyzing regions of dynamic stability in linear segments of power systems within the reduced ODE/PDE approach. We anticipate using this approach to extend the probabilistic distance to failure methods of \cite{11CPM,11CSPB} based on large deviation techniques to account for dynamic failures associated with the voltage collapse/instability.

\item Our ODE approach can be extended to analyze two-dimensional regions of power grids,  such as coarse-grained models of densely populated areas of the Eastern Interconnection in US.

\item The newly developed stochastic ODE/PDE technique will be an instrument for developing new control schemes based on detection of dangerous stochastic contingencies (see aforementioned discussion of the distance to failure approach) before they actually happen and mitigating them efficiently by using a variety of available energy and power resources.

\item We plan to use this ODE/PDE approach to study other types of instabilities in power grids, in particular these associated with synchronization-desynchronization transitions in phase dynamics \cite{10DB}, effects of inverse cascade (broadening of the desynchronization frequency spectra) \cite{11SM}, and electro-mechanical waves \cite{Parashar2004}.

\item It will also be important to extend the ODE/PDE model-reduction ideas to extracting/learning static and dynamic model parameters (especially these related to small scale distributed generation and loads) from measurements, in particular coming from the emergent Phasor Measurement Unit (PMU) technology \cite{PMU}.
\end{itemize}

%{\color{red} !!!!! Acknowledgments are needed only for the final submission - hide before submitting}
\section{Acknowledgments}

We are thankful to the participants of the "Optimization and
Control for Smart Grids" LDRD DR project at Los Alamos
and Smart Grid Seminar Series at CNLS/LANL for multiple
fruitful discussions. The work at LANL was carried out under the auspices of the National
Nuclear Security Administration of the U.S. Department of Energy at Los
Alamos National Laboratory under Contract No. DE-AC52-06NA25396.
The work of MC and SB was funded in part by DTRA/DOD under the grant
BRCALL06-Per3-D-2-0022 on "Network Adaptability from WMD Disruption
and Cascading Failures".

\appendix
\section{DistFlow Representation}
\label{sec:DistFlow}

In the case of a linear segment it may be convenient to restate the power flow equations in terms of the power flowing at any point of the segment.  In accordance with the original (spatially discrete) formulation of Eq.~(\ref{PF_theta},\ref{PF_v}), one establishes the following relation between the power leaving a site $n$ and moving towards the next site along the line, $n+1$,  and respective characteristics of the potential:
\begin{eqnarray}
&& P_n=%\mbox{Re}\left(V_n\frac{V_{n}^*-V_{n+1}^*}{R-i X}\right)
\frac{X}{R^2+X^2} v_n v_{n+1}\sin(\theta_n-\theta_{n+1})\nonumber\\ &&
+\frac{R}{R^2+X^2} v_n(v_n-v_{n+1}\cos(\theta_{n+1}-\theta_n)),
\label{Pn}\\
&&  Q_n=%\mbox{Im}\left(V_n\frac{V_{n}^*-V_{n+1}^*}{R-i X}\right)=
\frac{X}{R^2+X^2} v_n(v_n-v_{n+1}\cos(\theta_n-\theta_{n+1}))\nonumber\\ && -
\frac{R}{R^2+X^2} v_n v_{n+1}\sin(\theta_n-\theta_{n+1}).
\label{Qn}
\end{eqnarray}
In the continuous limit these relations become Eqs.~(\ref{PQ_cont}),  while Eqs.~(\ref{PF_theta_cont},\ref{PF_v_cont}) turn into
\begin{eqnarray}
0=p-\partial_z P-r\frac{P^2+Q^2}{v^2},\quad
0=q-\partial_z Q-x\frac{P^2+Q^2}{v^2}, \label{QP_cont}
\end{eqnarray}
where $r=R/a$, $x=X/a$ are resistance and inductance densities, and, in accordance with Eq.~(\ref{PQ_cont}), $P$ and $Q$ are related to $v$ as follows
\begin{eqnarray}
\partial_z v=-\frac{r P+ x Q}{v}. \label{v_eq}
\end{eqnarray}
In their discrete version the set of Eqs.~(\ref{QP_cont},\ref{v_eq}) for a line are called DistFlow equations \cite{89BWa,89BWb}. (See also \cite{10TSBCa,10TSBCb,11TSBC}.)

{\small
\bibliographystyle{unsrt}
\bibliography{voltage}

\begin{thebibliography}{10}

\bibitem{98AL}
V.~Ajjarapu and B.~Lee.
\newblock Bibliography on voltage stability.
\newblock {\em Power Systems, IEEE Transactions on}, 13(1):115 --125, feb 1998.

\bibitem{VanCutsem2000}
T.~Van~Cutsem.
\newblock Voltage instability: phenomena, countermeasures, and analysis
  methods.
\newblock {\em Proceedings of the IEEE}, 88(2):208 --227, feb 2000.

\bibitem{Weedy1968}
B.M. Weedy and B.R. Cox.
\newblock Voltage stability of radial power links.
\newblock {\em Electrical Engineers, Proceedings of the Institution of},
  115(4):528 --536, april 1968.

\bibitem{Venikov1975}
V.A. Venikov, V.A. Stroev, V.I. Idelchick, and V.I. Tarasov.
\newblock Estimation of electrical power system steady-state stability in load
  flow calculations.
\newblock {\em Power Apparatus and Systems, IEEE Transactions on}, 94(3):1034
  -- 1041, may 1975.

\bibitem{Venikov1977}
V.~A. Venikov.
\newblock {\em Transient Processes in Electrical Power Systems}.
\newblock English Translation, MIR Publishers, Moscow, 1977.

\bibitem{Taylor1994}
C.W. Taylor.
\newblock {\em Power System Voltage Stability}.
\newblock McGraw-Hill Inc., 1994.

\bibitem{Cutsem1998}
Thierry~van Cutsem and Costas Vournas.
\newblock {\em Voltage Stability of Electric Power Systems}.
\newblock Springer, 1998.

\bibitem{Kundur1993}
P.~Kundur.
\newblock {\em Power System Stability and Control}.
\newblock McGraw-Hill Inc., 1993.

\bibitem{Parashar2004}
M.~Parashar, J.S. Thorp, and C.E. Seyler.
\newblock Continuum modeling of electromechanical dynamics in large-scale power
  systems.
\newblock {\em Circuits and Systems I: Regular Papers, IEEE Transactions on},
  51(9):1848 -- 1858, sept. 2004.

\bibitem{Thorp1998}
J.S. Thorp, C.E. Seyler, and A.G. Phadke.
\newblock Electromechanical wave propagation in large electric power systems.
\newblock {\em Circuits and Systems I: Fundamental Theory and Applications,
  IEEE Transactions on}, 45(6):614 --622, jun 1998.

\bibitem{DeMarco1987}
C.~De~Marco and A.~Bergen.
\newblock A security measure for random load disturbances in nonlinear power
  system models.
\newblock {\em Circuits and Systems, IEEE Transactions on}, 34(12):1546 --
  1557, dec 1987.

\bibitem{Dobson1989a}
I~Dobson and H.-D. Chiang.
\newblock {Towards a theory of voltage collapse in electric power systems}.
\newblock {\em Systems \& Control Letters}, 13(3):253--262, September 1989.

\bibitem{Hill1993}
D.J. Hill.
\newblock {Nonlinear dynamic load models with recovery for voltage stability
  studies}.
\newblock {\em IEEE Transactions on Power Systems}, 8(1):166--176, 1993.

\bibitem{Davy1997}
R.J. Davy and I.a. Hiskens.
\newblock {Lyapunov functions for multimachine power systems with dynamic
  loads}.
\newblock {\em IEEE Transactions on Circuits and Systems I: Fundamental Theory
  and Applications}, 44(9):796--812, 1997.

\bibitem{Lesieutre2005}
BC~Lesieutre.
\newblock {Improving Dynamic Load and Generator Response Performance Tools}.
\newblock Technical report, Lawrence Berkeley Natioanal Laboratory, 2005.

\bibitem{PNNL-16916}
N~Lu and A~Qiao.
\newblock {Composite Load Model Evaluation}.
\newblock Technical Report September, PNNL, 2007.

\bibitem{WECC}
Western Electricity Coordinating Council (WECC).
\newblock {\em \url{http://www.wecc.biz}}.

\bibitem{PSLF}
Positive Sequence Load Flow Software (PSLF).
\newblock {\em
  \url{http://site.ge-energy.com/prod_serv/products/utility_software/en/ge_psl%
f/index.htm}}.

\bibitem{11CPM}
M.~Chertkov, F.~Pan, and M.G. Stepanov.
\newblock Predicting failures in power grids: The case of static overloads.
\newblock {\em Smart Grid, IEEE Transactions on}, 2(1):162 --172, march 2011.

\bibitem{11CSPB}
M.~Chertkov, M.G. Stepanov, F.~Pan, and R.~Baldick.
\newblock Exact and efficient algorithm to discover extreme stochastic events
  in wind generation over transmission power grids.
\newblock {\em invited session on Smart Grid Integration of Renewable Energy:
  Failure analysis, Microgrids, and Estimation at CDC/ECC 2011}, abs/1104.0183,
  2011.

\bibitem{10DB}
F.~D\"{o}rfler and F.~Bullo.
\newblock Synchronization and transient stability in power networks and
  non-uniform kuramoto oscillators.
\newblock In {\em American Control Conference (ACC), 2010}, pages 930 --937, 30
  2010-july 2 2010.

\bibitem{11SM}
Y.~Susuki and I.~Mezic.
\newblock Nonlinear koopman modes and coherency identification of coupled swing
  dynamics.
\newblock {\em Power Systems, IEEE Transactions on}, PP(99):1, 2011.

\bibitem{PMU}
Phasor Measurement Unit (PMU) technology.
\newblock {\em \url{http://en.wikipedia.org/wiki/Phasor_measurement_unit}}.

\bibitem{89BWa}
M.~Baran and F.F. Wu.
\newblock Optimal sizing of capacitors placed on a radial distribution system.
\newblock {\em Power Delivery, IEEE Transactions on}, 4(1):735 --743, jan 1989.

\bibitem{89BWb}
M.E. Baran and F.F. Wu.
\newblock Optimal capacitor placement on radial distribution systems.
\newblock {\em Power Delivery, IEEE Transactions on}, 4(1):725 --734, jan 1989.

\bibitem{10TSBCa}
K.~Turitsyn, P.~Sulc, S.~Backhaus, and M.~Chertkov.
\newblock Distributed control of reactive power flow in a radial distribution
  circuit with high photovoltaic penetration.
\newblock In {\em Power and Energy Society General Meeting, 2010 IEEE}, pages 1
  --6, july 2010.

\bibitem{10TSBCb}
K.~Turitsyn, P.~Sulc, S.~Backhaus, and M.~Chertkov.
\newblock Local control of reactive power by distributed photovoltaic
  generators.
\newblock In {\em Smart Grid Communications (SmartGridComm), 2010 First IEEE
  International Conference on}, pages 79 --84, oct. 2010.

\bibitem{11TSBC}
K.~Turitsyn, P.~Sulc, S.~Backhaus, and M.~Chertkov.
\newblock Options for control of reactive power by distributed photovoltaic
  generators.
\newblock {\em Proceedings of the IEEE}, 99(6):1063 --1073, june 2011.

\end{thebibliography}
}

\end{document}